\documentclass[sigconf,natbib=true]{acmart} 
\AtBeginDocument{%
  \providecommand\BibTeX{{%
    \normalfont B\kern-0.5em{\scshape i\kern-0.25em b}\kern-0.8em\TeX}}}

\usepackage{color}
\usepackage{algorithm}
\usepackage{algorithmic}
\usepackage{graphicx}
\usepackage{amsmath}
\usepackage{multirow}
\usepackage{balance} 

\copyrightyear{2022}
\acmYear{2022}
\setcopyright{acmcopyright}\acmConference[CIKM '22]{Proceedings of the 31st ACM
International Conference on Information and Knowledge Management}{October
17--21, 2022}{Atlanta, GA, USA}
\acmBooktitle{Proceedings of the 31st ACM International Conference on Information
and Knowledge Management (CIKM '22), October 17--21, 2022, Atlanta, GA, USA}
\acmPrice{15.00}
\acmDOI{10.1145/3511808.3557618}
\acmISBN{978-1-4503-9236-5/22/10}

\begin{document}

\title{Intra-session Context-aware Feed Recommendation in Live Systems}

\author{Luo Ji}
\affiliation{%
  \institution{DAMO Academy, Alibaba Group}
  \city{Hangzhou}
  \country{China}
}
\email{jiluo.lj@alibaba-inc.com}

\author{Gao Liu}
\affiliation{%
  \institution{DAMO Academy, Alibaba Group}
  \city{Hangzhou}
  \country{China}
}
\email{liugao.lg@alibaba-inc.com}

\author{Mingyang Yin}
\affiliation{%
  \institution{DAMO Academy, Alibaba Group}
  \city{Hangzhou}
  \country{China}
}
\email{hengyang.ymy@alibaba-inc.com}

\author{Hongxia Yang}\authornote{Corresponding author}
\affiliation{%
  \institution{DAMO Academy, Alibaba Group}
  \city{Hangzhou}
  \country{China}
}
\email{yang.yhx@alibaba-inc.com}

\renewcommand{\shortauthors}{Luo Ji, Gao Liu, Mingyang Yin, \& Hongxia Yang}
\begin{abstract}
Feed recommendation allows users to constantly browse items until feel uninterested and leave the session, which differs from traditional recommendation scenarios. Within a session, user's decision to continue browsing or not substantially affects occurrences of later clicks. However, such type of exposure bias is generally ignored or not explicitly modeled in most feed recommendation studies. In this paper, we model this effect as part of intra-session context, and propose a novel intra-session Context-aware Feed Recommendation (INSCAFER) framework to maximize the total views and total clicks simultaneously. User click and browsing decisions are jointly learned by a multi-task setting, and the intra-session context is encoded by the session-wise exposed item sequence. We deploy our model online with all key business benchmarks improved. Our method sheds some lights on feed recommendation studies which aim to optimize session-level click and view metrics.


\end{abstract}



\maketitle

\section{Introduction}

In recent years, feed recommendation (FR) has gained increasing popularity by providing never-ending and content-blended feeds in a waterfall form of item exhibitions. Generally, ranking in FR shares similar methodology with traditional learning-to-rank (LTR) methods, including traditional pointwise methods, as well as pairwise and listwise \cite{cao2007ListNet,Burges2010LambdaMart} methods which consider surrounding effect around items. In the original configuration of modeling, it is assumed that user observes item candidates with equal probabilities. This assumption has been questioned by some eye-tracking studies and works have been done to reimburse the positional, exposure or selection bias \cite{chen2021bias}. Recently there are also increasing efforts to make the model more consistent with the actual live environments, including session-based recommendation \cite{Sheu2020CAGE, Huang2021MTD}, sequential recommendation \cite{Wang2019SRS} and context-aware recommendation \cite{Chang2021ConTDM}. Within the scope of these works, both inter- and intra- context impacts are thoroughly studied and components such as RNN, GNN, transformer are widely adopted to capture the context-aware user preferences.

However, these studies seldom consider the interactive scenario of user browsing decisions, even for algorithms designed especially for FR \cite{Wu2021FeedRec, Huang2021SSD}. In a typical waterfall form of feeds, people first see the top item in default, then decide if click it or browse the next item. Obviously, this browsing decision is different from the click decision, and the next item could not be observed without that browsing operation \footnote{In our definition, a user click itself is not considered as the end of session, but it is the end if the user never returned from the clicked page.}. As a result, the likelihood of a user's later click is affected by previous browsing behavior, which results in substantial exposure bias. Furthermore, this time dependency on different item positions suggests that global views or clicks might be better business metrics than click-through rate (CTR), to highlight the user stickiness. Given such an objective, optimizing the instant CTR could reach localized optima which is deviated from the global optimum (for example, putting the most favorable item on the top might not always be a statistically best idea, since user would probably click it then leave the session immediately.). Unfortunately, most FR models by so far neither do not have explicit modeling of the browsing behaviors, nor try to solve the global metrics. While some optimization-based method such as reinforcement learning (RL) could be theoretically suitable to solve such type of problems, they are usually subject to computational complexity or exploration cost thus industrial application is limited. In this work, we try to build a supervised framework in which user sequential behaviors are explicitly modeled as a series of Markov events of clicks and scrolls, and total views and clicks are ranking objectives. This is to the best of our knowledge the first time to eliminate the session exposure bias of FR by such a methodology.

In this paper, we propose a novel \textbf{IN}tra-\textbf{S}ession \textbf{C}ontext-\textbf{A}ware \textbf{FE}ed \textbf{R}ecommendation (INSCAFER) framework which considers the aforementioned mechanism on mobile-based applications. On the mobile devices, browsing items are achieved by user operations of scrolling the screen down. To model user scroll and click behaviors, we define the intra-session context as the user browsing experience within the current session. We model the browsing and clicking events within a session as a Markov Chain, with the intra-session context as a latent variable, and click and scroll decisions conditioned on it. We solve the problem by maximizing the negative likelihood loss of the session events sequence, which is converted into a multi-task classification training with historical click and scroll as ground truth labels. The entire framework is similar to Generative Pre-Train (GPT) \cite{Radford2018GPT}, including a pre-training stage and a sequence generation stage. During pre-training stage, an intra-session context encoder is co-trained with a long-term interest net and a Multi-gate Mixture-of-Experts (MMOE) \cite{Ma2018MMOE} module. The pre-trained encoder is then deployed on the server, and a recommendation sequence generation task is conducted with intra-session context dynamically calculated during servicing stage. The major contributions of this paper are as follows:

\begin{itemize}

\item We explicitly consider the user scroll decisions and subsequent time-dependency of intra-session behaviors, which is more close to the real FR scenario. 
\item The model loss and architecture are designed from a theoretical starting point, with the objective of the expected total views and clicks.



\item We design a clear and fast framework similar to GPT to train and launch the model in a large-scale industrial system. 


\end{itemize}

\vspace{-5mm}
\begin{figure}[h]
  \centering
  \includegraphics[width=9cm]{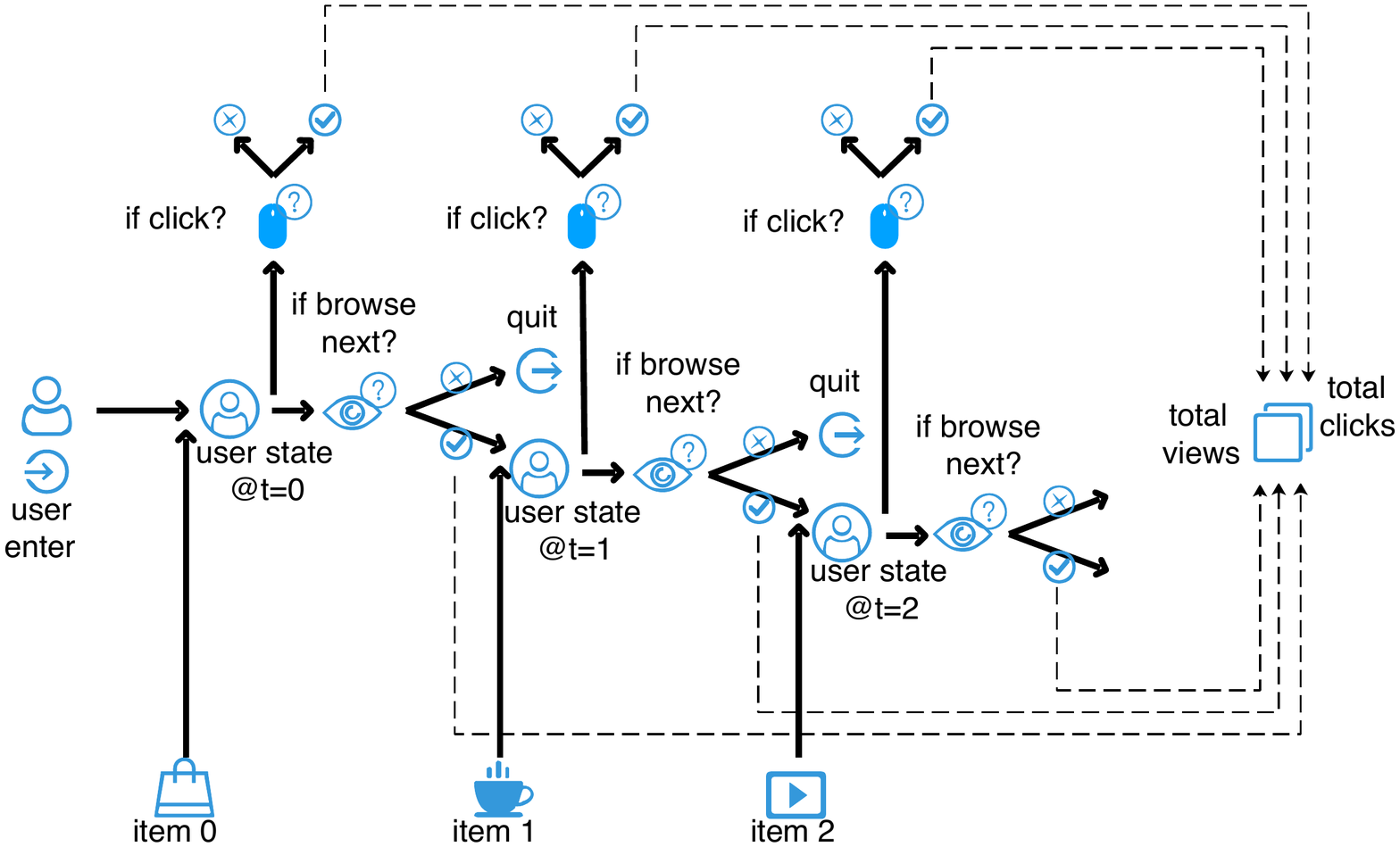} 
  \caption{The logical flow of user events in a typical feed session. At each position $t$, user decides if click the current item and/or browse the next item, or quit the session. The session-level recommending objective is to maximize the total views and clicks.}
  \label{feed_session_events}
\end{figure}

\section{Method}
\label{sec:method}




\subsection{Markov Modeling of Session Events}
\label{subsec:ubm}

The basic idea of our work is motivated from the concept of session events shown in Figure \ref{feed_session_events}, which characterizes typical FR scenario from traditional recommendation. User views the first item by default when entering into the session. Upon the $t$th item is $seen$, user makes two independent decisions: $click$ indicating whether click the current item, and $scroll$ indicating whether scrolling the screen down to browse the next item. The session stops when a 'not scroll any more' decision is made. At a specific position, the probability of click and scroll can be expressed by multiplication of two conditional probabilities
\begin{equation}
\label{eq:pos_event}
P(scroll, click \vert seen)P(seen \vert user, item, position)
\end{equation}
where ‘item’ is the item profile, and ‘user’ denotes the user perception affected by both the user long-term interest as well as the short-term experience within the current session. From this manner of definition, $user$ should be a latent variable recurrently affected by previous events. As a result, expressions of Eq. \ref{eq:pos_event} at different positions are not independent and identically distributed (i.i.d.) (\textit{e.g.}, a false $click$ will stops the session such that later $seen$ is always false), which violates the basic assumption of pointwise LTR. Instead, Figure \ref{feed_session_events} indicates an unidirectional time-dependency of timely event sequence, \textit{i.e.}, the former decisions will impact the later decisions, but reversely not. Similar conception can be found from the 'user cascade model' in \cite{Dupretg2008UBM} and also in DIEN \cite{Zhang2019DLEN}. On the contrary, bidirectional time-dependent methods such as \cite{Sun2019bert4rec} might be more suitable for Top-K recommendation instead of FR. 



Here we introduce some abbreviated notations of events, in which the latent variable $h$ denotes $user$, $x$ represents the item embedding, $c$ denotes probability of $click$ and $s$ denotes probability of $scroll$. $T$ is the total length of session and a specific position is $t \in [0, T]$. The bold version of variable denotes a sequence of events in a timely order, \textit{e.x.} $\mathbf{c} := \{c_0, c_1, \cdots, c_T \}$. Given an ordered exhibition of items $\mathbf{x}$, the joint probability of feed session events is
\begin{align}
   P(\mathbf{c}, \mathbf{s}, \mathbf{h} | \mathbf{x}) &= \prod_{t=0}^{T} P(c_t|h_t, x_t) P(s_{t}|h_t, x_t)P(h_{t+1}|h_t, x_t, s_{t-1}) \label{eq:session_seq_model} 
\end{align}
considering the Markov dependency. The objective is then to maximize the expectation of total views and clicks of all sessions
\begin{equation}
\label{eq:V}
    \max V = \mathbb{E}\sum_{t=0}^T (c_t + s_t)
\end{equation}



\subsection{Model}
\label{subsec:model}

\begin{figure}
  \centering
  \includegraphics[width=10cm]{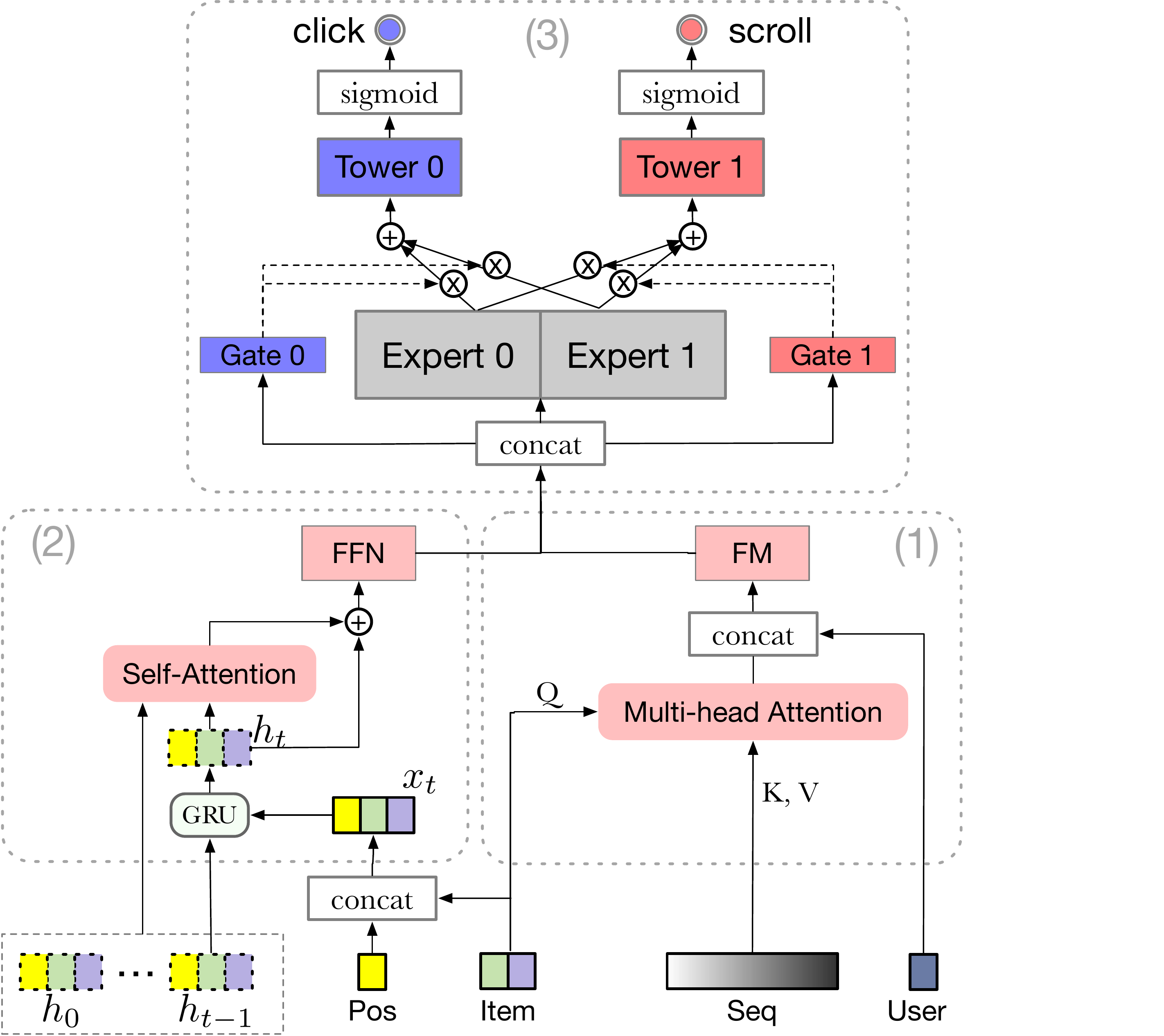} 
  \caption{Network structure for logits calculation at each ranking position, including three components: (1) The Interest net, (2) The context encoder, and (3) The MMOE module. More configuration details can be referred to Subsection \ref{sec:config}.}
  \label{network_structure}
  \vspace{-3.5mm}
\end{figure}

Similar with traditional LTR, optimizing the objective in Eq. (\ref{eq:V}) can be derived from maximizing the likelihood of Eq. (\ref{eq:session_seq_model}), by parameterizing the conditional probabilities by trainable $\theta$
\begin{equation}
\label{eq:mle}
\begin{split}
   \min_{\mathbf{h}, \theta} L(\mathbf{h}; \theta) = & -\text{log} P_{\theta}(\mathbf{c}, \mathbf{s} | \mathbf{h}, \mathbf{x}) = -\text{log} \frac{P_{\theta}(\mathbf{c}, \mathbf{s}, \mathbf{h} | \mathbf{x})}{P_{\theta}(\mathbf{h} | \mathbf{x})} \\
= & -\text{log} \prod_{t=0}^{T} P_{\theta}(c_t|h_t, x_t) P_{\theta}(s_{t}|h_t, x_t) \\
= & - \sum_{t=0}^{T} \text{log} P_{\theta}(c_t|h_t, x_t) + \text{log} P_{\theta}(s_t|h_t, x_t) 
\end{split}
\end{equation}
which indicates that the session-wise learning can be decomposed into consecutive position-wise multitask learning, jointly with estimation of the recurrent contextual state $h$. At each position, learning is multitask with BCEs of two classifications, $click$ and $scroll$. Although derivation of Eq. \eqref{eq:mle} seems trivial, there is no previous work which explicitly model the scroll behavior aside with click as multitask problem, to the best of our knowledge.


Eq. \eqref{eq:mle} implies our model structure, as shown in Figure \ref{network_structure}, with three modules: 1) the Interest net with a multi-head attention block followed by a Factorization Machines (FM) \cite{Rendle2010FM} layer which studies the user long-term interest; 2) the intra-session context encoder which models user short-term interest shift, mutual-item influence and the positional impact encoded by GRU, followed by a self-attention \cite{vaswani2017attention} block; and 3) an MMOE \cite{Ma2018MMOE} module with $click$ and $scroll$ subtasks. We name our approach as intra-session context-aware feed recommendation (INSCAFER).



\subsection{Learning and service framework}
\label{subsec:framework}

\begin{figure}
  \centering
  \includegraphics[width=8cm]{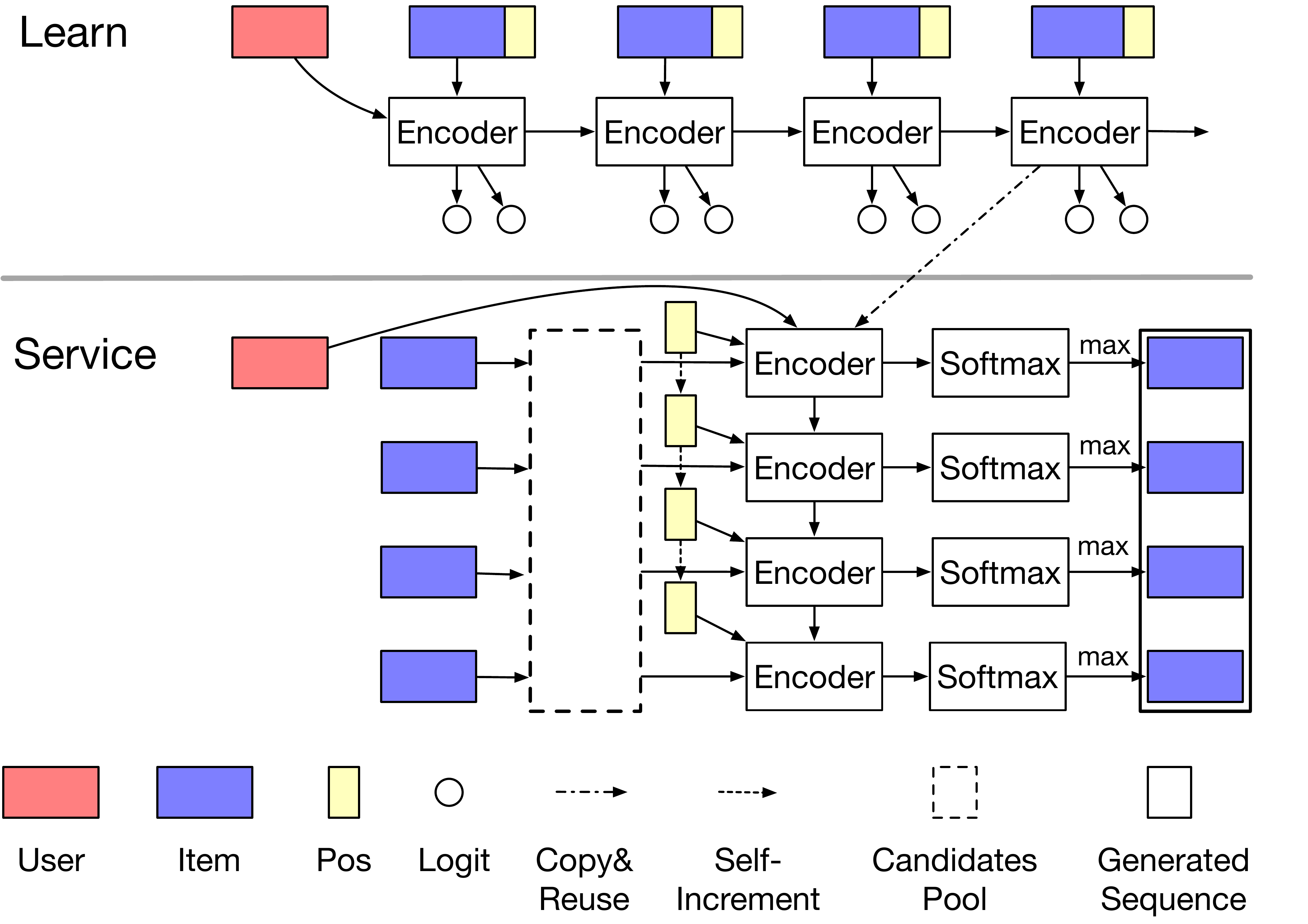} 
  \caption{Framework of INSCAFER. Learn: Sequential context encoding. Service: Greedy sequence generation.} 
  \label{learn_predict_framework}
  \vspace{-3.5mm}
\end{figure}

INSCAFER has a similar learning paradigm with the famous Generative Pre-Train (GPT) framework \cite{Radford2018GPT}, which is decoupled into two stages: the offline pre-training and online sequence generation, as shown in Figure \ref{learn_predict_framework}. During the offline training stage, the context encoder is first initialized with the user embedding, then recurrently updated with each exposed item embedding concatenated with the position lookup embedding as input;  while the MMOE logits of click ($\text{logit}^{\text{c}}$) and scroll ($\text{logit}^{\text{s}}$) are supervised by their labels \footnote{More precisely, this stage is similar with the 'supervised fine-tuning task' of GPT.}. During the service stage, a greedy sequence generation task is executed with the pre-trained context encoder retrieved and inferenced. The position embedding is self-incremented and looked up during this stage. With $K$ items left, the next item is decided by maximizing the following softmax:
\begin{equation}
\text{arg}\max_k \frac{\exp{(\text{logit}^{\text{c}}_k + \text{logit}^{\text{s}}_k)}}{\sum_{k^{\prime}}^K \exp{(\text{logit}^{\text{c}}_{k^{\prime}} + \text{logit}^{\text{s}}_{k^{\prime}})}}
\end{equation}


\section{Experiment}
\label{sec:experiment}

We apply our approach on a world-leading E-Payment platform which also provides a comprehensive recommendation application, including goods, restaurants and online services. We first perform substantial offline evaluations, including the classification metrics of click and scroll, to prove the superiority of INSCAFER. Ablation study is also conducted to verify the necessary of different components. Data \footnote{https://tianchi.aliyun.com/dataset/dataDetail?dataId=109858\&lang=en-us} and codes \footnote{https://github.com/AaronJi/RecINSCAFER} have been made public.



\subsection{Configurations}
\label{sec:config}

\begin{figure}
  \centering
  \includegraphics[width=9cm]{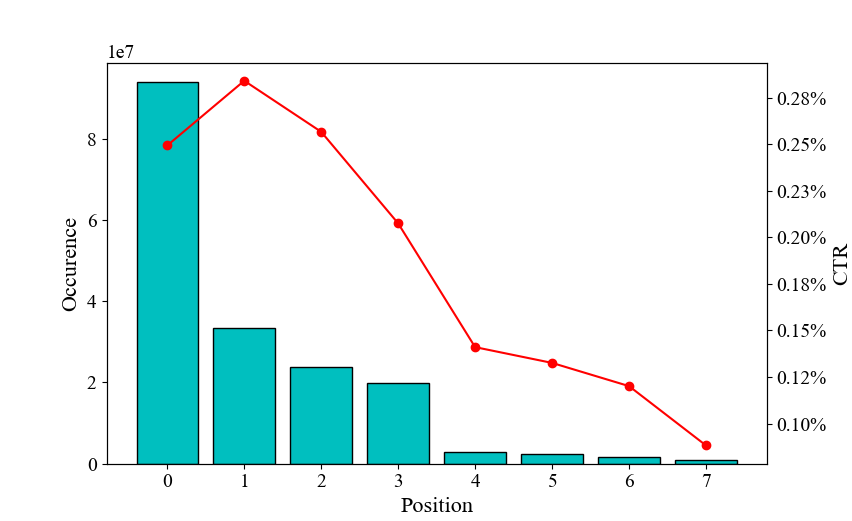} 
  \caption{View occurrences and CTR curves w.r.t. positions. Data is slightly rescaled due to confidential requirement.}
  \label{position_stat}
  \vspace{-3.5mm}
\end{figure}

Figure \ref{position_stat} shows the distribution of view occurrences and CTR according to the top 8 positions. One can see both of them decay quickly as position becomes larger \footnote{Note CTR at the first position is lower than the second and the third ones. The reason is that it is by design user will observe the first item by default. Only active and interested user will scroll down the screen and see the next items. Our model is able to capture such effects by explicit modeling.}, resulting in tremendous exposure bias for long sessions. Significant deviation would be introduced with an i.i.d. assumption.


System has about 200 million users and a billion number of views per day. Upon each query at most $15$ items are selected and sorted for exhibition. Embedding dimension is set to 16. Two heads are encoded in the multi-head attention layer and the cosine form of similarity is used. The shape of GRU latent state is the same with item embedding. There are 8 experts, 4 tasks and 64 as the expert shape in the MMOE module. The tower nets in MMOE are MLP with [128, 32] hidden units and sigmoid as the last activation. The ADAM optimizer is used with learning rate of 0.0001. Training costs more than 180k steps and loss converges after about 60k steps. 

\subsection{Offline Evaluation}
\label{subsec:offline_test}




Classification metric such as Area Under the ROC Curve (AUC) is evaluated for both $click$ and $scroll$ tasks. Experiments are repeated 10 times to report the averaged metric. Considering that our main purpose is to validate the effectiveness of loss in Eq. \eqref{eq:mle}, we choose offline baselines which are close to our online model, as well as some combinations of them and modules of \textit{INSCAFER}:

\begin{itemize}
    \item \textit{DIN}: the Deep Interest Network \cite{Zhou2018DIN}, as a standardized solution of industrial pointwise CTR model..
    \item \textit{DIN-ListNet}: A listwise version of Vanilla with loss replace by a listwise loss defined in \cite{cao2007ListNet}.
    \item \textit{MMOE}: An MMOE model with $click$ and $scroll$ as subtasks, sharing the same bottom structure with \textit{INSCAFER} \cite{Ma2018MMOE}.
    \item \textit{GRU4rec}: A session-based method with a latent state calculated by GRU \cite{Hidasi2016GRUrec}.
    \item \textit{GRU4rec-MMOE}: The \textit{GRU4rec} method combined with the MMOE structure to include $scroll$ consideration.
    \item \textit{Ptr-net}: The pointer network method \cite{vinyals2015ptrnet} which applies the seq2seq idea combined with attention blocks, and has similar sequence generation mechanism with us. We train it in the supervised mode of $click$ like work in \cite{Bello2019Seq2Slate}.
\end{itemize}


We summarize these results in Table \ref{tab:offline_result}, with evaluation of the click-through rate abbreviated as ctr and scroll rate abbreviated as scr. One can found that our INSCAFER has the best AUC for ctr and scr, and the third best MRR of scr. As comparisons, DIN and Ptr-net have reasonable ctr performances, but can not predict scr well, verifying that $scroll$ has a different distribution with $click$. Simply switching its loss to the listwise form as in DIN-ListNet, can not solve the problem either. GRU4rec with the context consideration has an improved ctr performance but scr is still bad. On the other hand, MMOE can have good scr prediction because of its multitask setting but at the cost of ctr performance degradation. Combining GRU4rec with MMOE has similar balanced ctr and scr performances.

\begin{table}[t]
\center
\caption{Comparison of Offline Performance} 
\label{tab:offline_result}
\vspace{-1mm}
\scalebox{0.85}{
\renewcommand{\arraystretch}{1}
\begin{tabular}{cccc}
\toprule
\multicolumn{1}{c}{Model}&
\multicolumn{1}{c}{AUC-ctr} &  \multicolumn{1}{c}{AUC-scr} \\ 
\hline
	DIN &$0.7771 \pm 0.0042$ &$0.5673 \pm 0.0032$ \\ 
	DIN-ListNet &$0.7362 \pm 0.0053$ &$0.5186 \pm 0.0061$ \\ 
	MMOE & $0.7682 \pm 0.0045$ &$0.7983 \pm 0.0048$ \\ 
	GRU4rec& $0.7812 \pm 0.0037$ &$0.5362 \pm 0.0049$  \\ 
	GRU4rec-MMOE& $0.7673 \pm 0.0044$ &$0.7885 \pm 0.0035$  \\ 
	Ptr-net& $0.7790 \pm 0.0065$ &$0.5642 \pm 0.0052$  \\ 
\hline
	INSCAFER (w/o gru) & $0.7701 \pm 0.0043$ &$0.8313 \pm 0.0040$ \\ 
	INSCAFER (w/o atten) & $0.7792 \pm 0.0032$ &$0.8429 \pm 0.0052$ \\ 
	INSCAFER (w/o pos) &$0.7722 \pm 0.0034$ &$0.7918 \pm 0.0044$ \\ 
    INSCAFER & $\textbf{0.7836} \pm 0.0035$ &$\textbf{0.8562} \pm 0.0056$\\  
\bottomrule
\end{tabular}
}
\vspace{-3.5mm}
\end{table}

In Table \ref{tab:offline_result}, we also perform some ablation tests, by each time excluding the GRU unit, the self-attention block, or the position embedding. Not surprisingly, INSCAFER still has the best performance, suggesting these key components are all crucial.

\subsection{Live Experiments}
\label{subsec:online_test}

The live experiment starts on September 3th, 2021 and lasts for about a week, in which its baseline is DIN ensembled with a conversion rate (CVR) model as well as a  model optimizing user views, corresponding to comprehensive business considerations. Due to limited online resource and business performance requirements, we only launch INSCAFER and DIN to compare with the baseline. Gains of some important business metrics of INSCAFER and DIN are shown in Table \ref{tab:live_result}. Not surprisingly, DIN can improve CTR performance further but at the cost of other metrics. In the contrary, our method have increased almost all key indicators, especially for views per user ($1.16\%$), scrolls per user ($1.56\%$), and total conversions ($1.56\%$). This solid live result indicates our model formulation has a better depiction of FR scenario and provides a more reasonable solution for global metric optimization.
%


\begin{table}[htb]
\center
\caption{Gains of Live Metrics to Online Baseline} 
\label{tab:live_result}
\vspace{-1mm}
\scalebox{1.0}{
\renewcommand{\arraystretch}{1}
\begin{tabular}{ccc}
\toprule
\multicolumn{1}{c}{Model}&
\multicolumn{1}{c}{DIN} & \multicolumn{1}{c}{INSCAFER} \\
\hline
	CTR & $\textbf{+0.20\%}$& $+0.13\%$ \\
	total views & $+0.01\%$ &$\textbf{+0.24\%}$\\
	views per user & $-0.14\%$ &$\textbf{+1.16\%}$\\
	scrolls per user & $-0.54\%$ &$\textbf{+1.56\%}$\\
	views per session & $-0.06\%$ &$\textbf{+0.38\%}$\\
	users to scroll & $-0.06\%$ &$\textbf{+0.42\%}$\\
	total conversions & $-0.06\%$ &$\textbf{+1.56\%}$\\

	\bottomrule
	\end{tabular}
}
\vspace{-3.5mm}
\end{table}

\section{Conclusion}
\label{sec:conclusion}

In this paper, we propose a novel INSCAFER method which solves the feed recommendation problem by considering the intra-session context. In feed recommendation, user's previous operations, especial decisions to browsing the next item or not, affect distribution of later operations, resulting in substantial exposure bias. We start our model formation from maximizing the likelihood of the joint probability of session events, with explicit consideration of the timely dependency of user intra-session behaviors. Such intra-session context is studied during training and considered in servicing by a GPT-like framework. Both offline and live experiments have verified our method's superiority over several popular baselines. 




\bibliographystyle{ACM-Reference-Format}
\balance
\bibliography{main}

\end{document}